\documentclass[prb,twocolumn,floats]{revtex4}
\usepackage{amsmath}
\usepackage{amssymb}
\usepackage{graphicx}
\usepackage{epstopdf}
\begin{document}
\title{Effects of Strong Correlations on the Zero Bias Anomaly in the Extended  Hubbard Model with Disorder}
\author{Yun Song$^1$, S. Bulut,$^2$, R. Wortis$^2$, and W. A. Atkinson$^{2,3}$}
\affiliation{$^1$Department of Physics, Beijing Normal University,
Beijing 100875, China \\ $^2$Department of Physics and Astronomy,
Trent University, 1600 West Bank Dr., Peterborough ON, K9J 7B8,
Canada\\
$^3$Center for Electronic Correlations and Magnetism, EP VI, 
          Universit\"at Augsburg, D-86135 Augsburg, Germany}
\date{\today}

\begin{abstract}
We study the effect of strong correlations on the zero bias anomaly
(ZBA) in disordered interacting systems.  We focus on the
two-dimensional extended Anderson-Hubbard model, which has both
on-site and nearest-neighbor interactions on a square lattice.  We use
a variation of dynamical mean field theory in which the diagonal
self-energy is solved self-consistently at each site on the lattice
for each realization of the randomly-distributed disorder potential.
Since the ZBA occurs in systems with both strong disorder and strong
interactions, we use a simplified atomic-limit approximation for the
diagonal inelastic self-energy that becomes exact in the
large-disorder limit.  The off-diagonal self-energy is treated within
the Hartree-Fock approximation. The validity of these approximations
is discussed in detail.  We find that strong correlations have a
significant effect on the ZBA at half filling, and enhance the Coulomb
gap when the interaction is finite-ranged.
\end{abstract}
\maketitle

\section{Introduction}
The term ``zero bias anomaly" (ZBA) refers either to a peak in, or a
suppression of, the density of states (DOS) at the Fermi energy.  In
disordered materials, a ZBA arises from the interplay between disorder
and interactions. Zero bias anomalies were originally predicted to
occur in strongly-disordered insulators by Efros and
Shklovskii\cite{Efros1975a,Efros1975b} (ES) and later in
weakly-disordered metals by Altshuler and Aronov\cite{Altshuler1985}
(AA).  In the limit of weak interactions and disorder, AA showed that
the exchange self-energy of the screened Coulomb interaction produces
a cusp-like minimum in the DOS at the Fermi energy.  In the limit of strong
disorder, ES showed that the classical Hartree self-energy of the
unscreened Coulomb interaction causes the DOS to vanish at the Fermi
energy.  Experiments have shown a smooth evolution between the AA and
ES limits as a function of disorder,\cite{Butko2000} and it appears
that the essential physics of the ZBA in conventional metals and
insulators is well understood.

In this work, we are interested in anomalies which have been observed
in a number of transition metal oxide materials, where the physics is
less well understood.  Transition metal oxides often exhibit
unconventional behavior because the physics of their valence band is
dominated by strong short-ranged interactions whose effects cannot
generally be explained by conventional theories of metals and
insulators.  Most notably, many transition metal oxides exhibit a Mott
transition when their valence band is half-filled.  In disorder-free
systems, the Mott transition\cite{Dagotto1994,Imada1998} occurs between
a gapless metallic state and a gapped insulating state, and is driven
by a strong intraorbital Coulomb interaction.
 
The Mott transition may occur as a function of any number of
parameters,\cite{Imada1998} such as temperature\cite{Morin1959} or
magnetic field,\cite{Hanasaki2006} but more commonly occurs in
transition-metal oxides as a result of chemical doping. A number of
experiments\cite{Sarma1998,Ino2004,Kim2006,Nakatsuji2004,Kim2005} have
found that chemical doping introduces sufficient disorder that there
is a regime between the Mott-insulating and gapless phases that is
characterized by a ZBA. This naturally raises the question of how the
strong electron-electron correlations that are prevalent in the
gapless phase near the Mott transition affect the physics of the ZBA.

The effect of strong correlations on the ZBA has received little
attention.\cite{discussed} In large part, this is due to the
difficulty of incorporating both strong correlations and disorder in a
manageable theory.  A number of calculations based on the unrestricted
Hartree-Fock approximation (HFA) have been used to study the phase
diagram of the disordered Hubbard
model.\cite{Milovanovic1989,Tusch1993,Heidarian2004a,Fazileh2006} In
these calculations, the disorder potential is treated exactly for
finite-sized systems, but the intraorbital Coulomb interaction is
treated at the mean-field level and therefore neglects strong
correlations.  Much of the recent progress has involved various
formulations of dynamical mean field theory (DMFT) to include disorder
at some level of approximation. Coherent-potential-like approximations
have been employed by a variety of
authors.\cite{Ulmke1995,Laad2001,Nikolic2003,Vollhardt2005,Balzer2005,Lombardo2006}
In these calculations, the local electron self-energy contains both
inelastic contributions from the interactions and elastic
contributions from the disorder-averaging process.  It is well known
that these kinds of disorder-averaging approximations capture many
features of the DOS but do not retain the nonlocal correlations
responsible for the ZBA and cannot, therefore, explain the experiments
cited above.\cite{Altshuler1985} An extension of DMFT, called
statistical DMFT, has been employed to study ensembles consisting of
Bethe lattices with random site energies.
\cite{Kotliar1997,Miranda2005} This represents an improvement over the
disorder-averaged approximations in that the results depend
nontrivially on the coordination number of the Bethe lattice.  Very
recently, the DMFT equations have been solved by us on a two
dimensional square lattice in a way which preserves spatial
correlations between sites.\cite{Song2007} The calculations employed a
simple atomic-limit approximation for the self-energy that, while
generally appropriate for the large-disorder limit, does not contain
the off-diagonal self-energies responsible for the DOS anomalies
observed in experiments.

In this work, we extend our earlier calculations to include a nearest
neighbour interaction $V$.  This interaction is treated at the
mean-field level, and serves two purposes in this work: First, it
represents finite range interactions which are present in real
materials. Second, the exchange self-energy coming from this
interaction plays a qualitatively similar role to the off-diagonal
self-energy that is missing in our treatment of the intraorbital
interaction $U$.  The intraorbital self-energy is calculated within a
Hubbard-I (HI) approximation which has the effect of suppressing double
occupancy of orbitals. Details of these calculations are outlined in
Sec.~\ref{method}.  Because the HI approximation is known to be
defficient in the disorder-free limit, we include a discussion in
Sec.~\ref{connection} showing that the approximation for the diagonal
self-energy is valid in the limit of large disorder and low
coordination number.  In Sec.~\ref{cpa}, we introduce a coherent
potential approximation (CPA) for the disordered Hubbard model which
is used as a point of comparison for the lattice DMFT calculations.
The results of numerical and analytical
calculations are presented in Sec.~\ref{results}. 

Our primary result is that the ZBA is strongly enhanced by
electron correlations near half-filling when the interaction is
finite-ranged.  The ZBA is predominantly of the ES type, in the sense
that the largest contribution is from the classical Hartree
interaction between localized charges on neighboring sites.
The enhancement of the ZBA at half-filling is due to strong
correlations which inhibit screening of the impurity potential: Near
the Mott transition, the absence of screening drives the system
towards the strongly localized limit where the physics of the Coulomb
gap is most important.  The magnitude of the ZBA is doping dependent,
and drops off away from half-filling.

\section{Calculations}
\label{calculations}
\subsection{Method}
\label{method}

 We study an extended Anderson-Hubbard model, also known as a
disordered $t$-$U$-$V$ model.  In this model, $t$ denotes the electron
kinetic energy, $U$ the intraorbital Coulomb interaction, and $V$ the
interaction between nearest-neighbor sites.  The $V$-term in the
Hamiltonian can be used to represent two distinct physical processes.
First, it represents the nonlocal Coulomb interaction that, while
neglected in the Hubbard model, is generally present in real
materials.  Second, as shown below, the exchange self-energy from the
nonlocal interaction is qualitatively similar to the exchange
self-energy that arises in a perturbative treatment of the Hubbard
model.  This effective interaction plays a crucial role in clean
low-dimensional
systems.\cite{Lieb1968,Hirsch1985,White1989,Otsuka2000} It is often
treated explicitly in the large-$U$ limit via an approximate mapping
of the Hubbard model onto the $t$-$J$ model, where $J$ is the strength
of the effective nonlocal interaction.
A major difference between the $t$-$J$ and extended Anderson-Hubbard
models is that double occupation of orbitals is completely suppressed
in the former whereas a finite fraction of sites will be doubly
occupied in the latter when the width of the disorder distribution is
larger than $U$.

   The extended Anderson-Hubbard model is
\begin{eqnarray}
\hat H &=& -t \sum_{\langle i,j\rangle,\sigma}  c^\dagger_{i\sigma} c_{j\sigma}
 + \frac{V}{2}\sum_{\langle i,j\rangle} \hat n_i \hat n_j  \nonumber \\
&& +\sum_i \left ( \epsilon_i \hat n_i + U  \hat n_{i\uparrow}\hat
n_{i\downarrow} \right ),
\end{eqnarray}
where $\langle i,j\rangle$ denotes nearest-neighbor lattice sites $i$
and $j$, $\hat n_{i\sigma} = c^\dagger_{i\sigma} c_{i\sigma}$, $\hat
n_i = \hat n_{i\uparrow} + \hat n_{i\downarrow}$, and parameters $t$,
$U$ and $V$ are the kinetic energy, the on-site Coulomb interaction,
and the nearest-neighbor interaction respectively. $\epsilon_i$ is the
site energy, which is box-distributed according to
$P(\epsilon_i)=W^{-1}\Theta(W/2-|\epsilon_i|)$, where $W$ is the width
of the disorder distribution and $\Theta(x)$ the step-function.

We treat the nearest-neighbor interaction at the mean-field level:
\begin{equation}
\frac{V}{2}
\hat n_i \hat n_j \approx
V
\left ( 
\hat n_i n_j - \sum_\sigma
c^\dagger_{i\sigma} c_{j\sigma} f_{ji} + f_{ij}^2 - \frac{n_in_j}{2}
\right )
\end{equation}
with $f_{ji} = \langle c^\dagger_{j\uparrow}c_{i\uparrow}\rangle =
\langle c^\dagger_{j\downarrow}c_{i\downarrow}\rangle$ in the
paramagnetic phase and $n_j =\langle \hat n_j \rangle$.  Both
$f_{ij}$ and $n_j$ are determined self-consistently.
The mean-field Hamiltonian, up to an additive constant, is:
\begin{equation}
\hat H = \sum_{\langle i,j\rangle \sigma} t_{ij}^\prime c^\dagger_{i\sigma} c_{j\sigma}
+\sum_i   \epsilon_i^\prime  \hat n_i
+ U \sum_i \hat n_{i\uparrow}\hat n_{j\downarrow}
\label{MFham}
\end{equation}
where $t_{ij}^\prime = -t - V f_{ij}$ and $ \epsilon_i^\prime =
\epsilon_i + V \sum_j n_j $, where the sum is over nearest neighbors
of $i$.

The approximate Hamiltonian (\ref{MFham}) is then solved using an
iterative lattice DMFT (LDMFT) method that captures the
strong-correlation physics of the intraorbital
interaction.\cite{others} On an $N$-site lattice, the single-particle
Green's function can be expressed as an $N\times N$ matrix in the
site-index:
\begin{equation}
{\bf G}(\omega) = [\omega{ \bf I} - {\bf t} - {\boldsymbol \epsilon} - {\bf
\Sigma}(\omega)]^{-1}
\label{G}
\end{equation}
with ${\bf I}$ the identity matrix, ${\bf t}$ the matrix of
renormalized hopping amplitudes $ t_{ij}^\prime$, ${\boldsymbol
\epsilon}$ the diagonal matrix of renormalized site energies
$\epsilon_i^\prime$ and ${\bf \Sigma}(\omega)$ the matrix of local
self-energies.  The self-energy $\Sigma_i(\omega)$ corresponds to the
inelastic self-energy $\Sigma(\omega)$ (the so-called ``impurity
self-energy") in standard DMFT, which is obtained by various
self-consistent impurity solvers.\cite{Kotliar1996} The iteration
cycle begins with the calculation of ${\bf G}(\omega)$ from
Eq.~(\ref{G}), and $n_{i\sigma}$ and $f_{ij}$ given by,
\begin{eqnarray}
n_{i\sigma} &=& -\frac{1}{\pi}\int_{-\infty}^{\varepsilon_F} d\omega \mathrm{Im} G_{ii}(\omega) \\
f_{ij} &=& -\frac{1}{\pi}\int_{-\infty}^{\varepsilon_F} d\omega \mathrm{Im}
G_{ji}(\omega).
\end{eqnarray}
from the previous iteration. For each site $i$, one defines a Weiss
mean field ${\cal G}^0_i(\omega) = [G_{ii}(\omega)^{-1} +
\Sigma_i(\omega)]^{-1}$ where $G_{ij}(\omega)$ are the matrix elements
of ${\bf G}(\omega)$. The HI approximation is the simplest improvement
over the HFA that generates both upper and lower Hubbard bands and, as
we discuss in the next section, it works well in the large-disorder
limit $W\gg t$.  In this approximation, ${\cal G}_i(\omega) =\left[
{\cal G}^0_i(\omega)^{-1} - \Sigma_{i}^{HI}(\omega) \right ]^{-1}$
where
 \begin{equation}
 \Sigma_{i}^{HI}(\omega) = U\frac{n_i}{2} + \frac{U^2\frac {n_i}{2} (1-\frac {n_i}{2})}{\omega - \epsilon_i^\prime -U(1-\frac {n_i}{2})},
\label{eq:SE}
 \end{equation}
and $n_i$ is self-consistently determined for each site.\cite{caveat}

We remark that  we can also express the Green's function as 
\begin{equation}
{\bf G}(\omega) = [\omega{ \bf I} - {\bf t_0} - {\boldsymbol \epsilon_0} - {\bf
\Sigma}(\omega)]^{-1}
\label{G0}
\end{equation}
where $\bf t_0$ and $\boldsymbol \epsilon_0$ are matrices of the
unrenormalized hopping amplitudeds and site energies.  In this case,
the self-energy is:
\begin{subequations}
\begin{equation}
\Sigma_{ii} = 
U \frac{n_i}{2} + V\sum_j n_j + \frac{\displaystyle U^2 \frac{n_i}{2} (1-\frac{n_i}{2})}{\displaystyle \omega-\epsilon_i - U (1-\frac{n_i}{2}) -  V\sum_j n_j}
\label{eq:sii3}
\end{equation}
and 
\begin{equation}
\Sigma_{ij} = -Vf_{ij}.
\end{equation}
\label{eq:Sapprox}
\end{subequations}
This form emphasizes the nonlocal nature of the self-energy.

\subsection{Validity of the Self-Energy}
\label{connection}
In this section, we discuss our treatment of the intraorbital
interaction in the renormalized Hamiltonian (\ref{MFham}).  We show
that Eq.~(\ref{eq:SE}) is a reasonable approximation for the local
self-energy in the large-disorder limit provided that nonlocal
effective interactions generated by $U$ can be absorbed into the
interaction $V$.

Our discussion is based on a two-site Anderson-Hubbard Hamiltonian,
ie.\ on Eq.~(\ref{MFham}) with two sites labelled ``1" and ``2".  This
Hamiltonian can, of course, be diagonalized exactly with relatively
little effort.  Here, we are interested in developing an
approximate treatment that is valid in the large disorder limit, and
which can be applied to the $N$-site problem.  Comparison to the exact
solution is used as a benchmark for the approximation.

\begin{figure}
\includegraphics[width=\columnwidth]{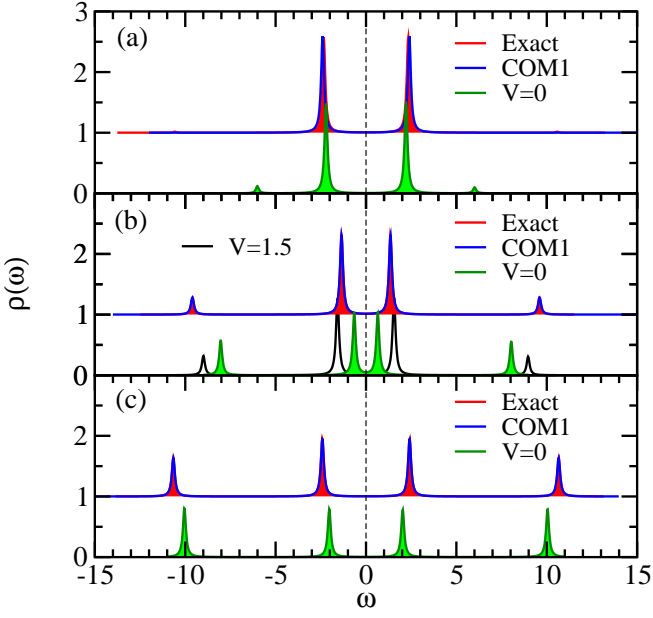}
\caption{(Color online) Comparison of the approximate and exact
densities of states for a two-site system.  The site energies are
$\epsilon_1 = W/2$, $\epsilon_2 = -W/2$ for $W = 8t$.  Figure panels
are for (a) $U=4t$, (b) $U=8t$ and (c) $U=12t$.  Exact calculations
are based on exact diagonalization of the two-site Hubbard model.
Approximate densities of states are given by $\rho(\omega) =
-\mbox{Im} \sum_i G_{ii}(\omega)/\pi$, with ${\bf G}(\omega)$ the
matrix (\protect\ref{eq:G3}).  Two different approximations are made
for the self energies: the first has the self energy given by
Eq.~(\protect\ref{eq:S}) and makes the COM1 approximation for
$p$;\protect\cite{Avella2004} the second has the self energy given by
Eq.~(\protect\ref{eq:Sapprox}) with different choices for $V$.  Both
the COM1 and exact solutions are offset for clarity.  The COM1
solution is essentially indistinguishable from the exact solution in
all cases.  Note that $\varepsilon_F$ is the zero of energy in this
and all other figures.  }
\label{fig:twosite}
\end{figure}

We use an equation-of-motion method to arrive at approximate
expressions for the single-particle Green's function.  Defining a
Liouvillian superoperator ${\cal L}$ such that\cite{Fulde}
\begin{equation}
{\cal L} \hat A \equiv [\hat H,\hat A],
\end{equation}
where $\hat A$ is an arbitrary operator, we can formally write the
time evolution of $\hat A$ as $\hat A(t) = \exp(i{\cal L} t) \hat
A(0)$.  It follows directly that the retarded Green's function can be
written
\begin{equation}
G_{i\sigma,j\sigma} (\omega) = ( c^\dagger_{i\sigma} | \frac{1}{\omega
- {\cal L}} c^\dagger_{j\sigma} )
\end{equation}
where the inner product of two operators is defined as $( \hat A |
\hat B) = \langle \{ \hat A^\dagger, \hat B\} \rangle$ and $\{,\}$
refers to the anticommutator.  The operator set $c_{i\sigma}^\dagger$
is not closed under operations by ${\cal L}$, but a complete operator
set can be generated with repeated operation by ${\cal L}$ on
$c_{i\sigma}^\dagger$.  For example,
\begin{equation}
{\cal L} c_{i\sigma}^\dagger = \epsilon_i^\prime c_{i\sigma}^\dagger+
\sum_j c_{j\sigma}^\dagger t_{ji}^\prime + U (b^\dagger_{i\sigma} -
c_{i\sigma}^\dagger n_{i\overline\sigma})
\end{equation}
where $b^\dagger_{i\sigma} = c_{i\sigma}^\dagger (\hat
n_{i\overline\sigma} - n_{i\overline\sigma})$ and $\overline\sigma =
-\sigma$.  The operator $b^\dagger_{i\sigma}$ is a composite operator,
and further composite operators can be generated from ${\cal L}^2
c_{i\sigma}^\dagger$, etc.  The higher order composite operators
involve excitations on multiple lattice sites, and are therefore
expected to be less important in the disordered case than in the clean
limit.  Here, we truncate the series after a single application of
${\cal L}$ so that our operator basis consists of two operators,
$c_{i\sigma}^\dagger$ and $b_{i\sigma}^\dagger$, for each site and
spin.  This leads to a ``two-pole" approximation for the Green's
function.  This approach has been studied at length in the clean limit
and has been shown to provide a reasonable qualitative description of
the Hubbard model.\cite{Mehlig1995,Grober2000,Avella2004} As shown in
Fig.~\ref{fig:twosite}, the two-pole approximation (described in more
detail below) is essentially indistinguishable from the exact solution
for the DOS of the two-site system.
 
It is useful to define a generalized Green's function in the expanded
operator space:
\begin{equation}
{\cal G}_{i\sigma,j\sigma} (\omega)  = 
\left [
\begin{array}{cc}
( c^\dagger_{i\sigma} | \frac{1}{\omega - {\cal L}} c^\dagger_{j\sigma} ) &
( c^\dagger_{i\sigma} | \frac{1}{\omega - {\cal L}} b^\dagger_{j\sigma} ) \\[3mm]
( b^\dagger_{i\sigma} | \frac{1}{\omega - {\cal L}} c^\dagger_{j\sigma} ) &
( b^\dagger_{i\sigma} | \frac{1}{\omega - {\cal L}} b^\dagger_{j\sigma} )
\end{array}
\right ]
\end{equation}
such that $G_{i\sigma,j\sigma} (\omega) $ is given by the upper left
quadrant of ${\cal G}_{i\sigma,j\sigma} (\omega) $.  Defining the
Liouvillian matrix,
\begin{equation}
L_{i\sigma,j\sigma} (\omega)  = 
\left [
\begin{array}{cc}
( c^\dagger_{i\sigma} |  {\cal L} c^\dagger_{j\sigma} ) &
( c^\dagger_{i\sigma} |  {\cal L} b^\dagger_{j\sigma} ) \\[2mm]
( b^\dagger_{i\sigma} | {\cal L} c^\dagger_{j\sigma} ) &
( b^\dagger_{i\sigma} | {\cal L} b^\dagger_{j\sigma} )
\end{array}
\right ],
\end{equation}
and the matrix of overlap integrals,
\begin{equation}
\chi_{i\sigma,j\sigma}   = 
\left [
\begin{array}{cc}
( c^\dagger_{i\sigma} |  c^\dagger_{j\sigma} ) &
( c^\dagger_{i\sigma} |  b^\dagger_{j\sigma} ) \\[2mm]
( b^\dagger_{i\sigma} |  c^\dagger_{j\sigma} ) &
( b^\dagger_{i\sigma} |  b^\dagger_{j\sigma} )
\end{array}
\right ],
\end{equation}
we get 
\begin{equation}
{\cal G}_{i\sigma,j\sigma} (\omega) = \sqrt{\chi_{i\sigma,i\sigma} }
[\omega - \tilde L]^{-1}_{i\sigma,j\sigma}
\sqrt{\chi_{j\sigma,j\sigma} }
\label{G2}
\end{equation}
with 
\begin{equation}
\chi_{i\sigma, j\sigma} = \delta_{i,j} \left [ 
\begin{array}{cc} 1 & 0 \\
 0 & n_{i\overline\sigma} (1-n_{i\overline\sigma}) 
 \end{array}
 \right ]
\end{equation}
 and $\tilde L =  \sqrt{\chi^{-1}} L \sqrt{\chi^{-1}}$.

For the nonmagnetic case, up and down spins are equivalent, and only
the former need be considered.  For the two-site system, and taking
the basis $[c_{1\uparrow}^\dagger, c_{2\uparrow}^\dagger,
b_{1\uparrow}^\dagger, b_{2\uparrow}^\dagger]$, we write the
Liouvillian matrix explicitly as
\begin{equation}
\tilde L  = 
\left [
\begin{array}{cccc}
\epsilon_1^\prime + Un_{1\downarrow} & -t^\prime & \tilde U_1 & 0 \\
-t^\prime & \epsilon_2^\prime + Un_{2\downarrow} & 0 & \tilde U_2 \\
\tilde U_1 & 0 & \tilde \epsilon_1 &  -t^\prime \tilde p \\
0 & \tilde U_2 & -t^\prime \tilde p & \tilde \epsilon_2
\end{array}
\right ],
\label{L}
\end{equation}
where 
\begin{eqnarray*}
\tilde U_i &=& U \sqrt{n_{i\downarrow} (1-n_{i\downarrow})} \\ 
\tilde \epsilon_i & = & \epsilon_i^\prime + U (1-n_{i\downarrow}) -
t^\prime \tilde \Delta_i \\
\tilde \Delta_i &=& \Delta_i[n_{i\downarrow}(1-n_{i\downarrow})]^{-1} \\
\Delta_i &=& \langle (\hat n_{i\uparrow} - n_{i\downarrow})
c^\dagger_{j\downarrow}c_{i\downarrow} \rangle - \langle (1-\hat
n_{i\uparrow} - n_{i\downarrow} ) c^\dagger_{i\downarrow}c_{j\downarrow}
\rangle \\
\tilde p & = & p [n_{1\downarrow}
(1-n_{1\downarrow})n_{2\downarrow} (1-n_{2\downarrow})]^{-1/2} \\
p &=& \langle \hat n_{1\downarrow}\hat n_{2\downarrow} \rangle -
n_{1\downarrow}n_{2\downarrow} - \langle c^\dagger_{1\uparrow}
c_{2\uparrow} ( c^\dagger_{2\downarrow} c_{1\downarrow} +
c^\dagger_{1\downarrow} c_{2\downarrow} ) \rangle
\end{eqnarray*}
In the expression for $\Delta_i$, $j$ is the nearest neighbor to $i$
(i.e.\ $j=1$ if $i=2$ and $j=2$ if $i=1$).

Equation [\ref{L}] allows us to solve for the Green's function,
provided the fields $\Delta_i$, $p$, and $n_{i\downarrow}$ are known.
In practice, $\Delta_i$ and $n_{i\downarrow}$ can be solved
self-consistently, but further information is needed to calculate $p$.

The Green's function $G(\omega)$ is given by the upper left $2\times
2$ quadrant of ${\cal G}(\omega)$.  It is straightforward to solve
Eq.~(\ref{G2}) to show that
\begin{equation}
G(\omega) = \left [ \begin{array}{cc}
\omega - \epsilon_1^\prime - \Sigma_{11}(\omega) &
t^\prime - \Sigma_{12}(\omega) \\
t^\prime - \Sigma_{21}(\omega) &
\omega - \epsilon_2^\prime - \Sigma_{22}(\omega) 
\end{array}
\right ]^{-1}
\label{eq:G3}
\end{equation}
with 
\begin{subequations}
\begin{eqnarray}
\Sigma_{ii} &=& U n_{i\downarrow} + \frac{\tilde U_i^2(\omega - \tilde \epsilon_j)}{(\omega-\tilde \epsilon_1)(\omega-\tilde \epsilon_2) - t^{\prime 2} \tilde p^2 } \label{sii} \\
\Sigma_{12} & = & \Sigma_{12} \nonumber \\
& = & \frac{-t^\prime U^2 p}{(\omega-\tilde \epsilon_1)(\omega-\tilde \epsilon_2) - t^{\prime 2} \tilde p^2 } \label{s12}
\end{eqnarray}
\label{eq:S}
\end{subequations}
where $j$ is again the nearest neighbor to $i$.  Equations [\ref{sii}]
and [\ref{s12}] are the basic expressions for the self-energy.  These
expressions are shown in Fig.~\ref{fig:twosite} to give
very accurate results for the Green's function provided that $p$ is
correctly chosen.  In this work, we have used the COM1
approximation of Avella and Mancini.\cite{Avella2004} The COM1 approximation
works well for the simple inhomogeneous systems we have tested it on, but
is extremely difficult to apply to disordered systems where
$p$ is different along every bond in the lattice. 

Some simplifications can be made in the case of large disorder.  To
begin with, we discuss the diagonal self-energies $\Sigma_{ii}$, and
take $i=1$.  We note that $\tilde p \sim O(1)$, so that
\begin{equation}
\Sigma_{11} \rightarrow U n_{1\downarrow} + \frac{U^2 n_{1\downarrow}
(1-n_{1\downarrow})}{\omega-\epsilon_1^\prime - U (1-n_{1\downarrow})
+ t^\prime \tilde \Delta_1},
\label{sii2}
\end{equation}
whenever $(\omega-\tilde \epsilon_1)(\omega-\tilde \epsilon_2) \gg
t^2$.  Apart from the shift $t\tilde \Delta_1$, this is just the
Hubbard-I approximation for the self-energy.  As we show next,
Eq.~[\ref{sii2}] is justified for $\omega\approx \varepsilon_F$ in the
large disorder limit, which corresponds in the two-site problem to
$|\epsilon_1 - \epsilon_2| \gg t$.

We are interested in the validity of Eq.~(\ref{sii2}) near
$\varepsilon_F$ and take the particular case $\varepsilon_F=U/2$
(which corresponds to half-filling) where strong correlations are most
important.  When $t=0$, each atomic Green's function has poles at
$\epsilon_i^\prime$ and $\epsilon_i^\prime+U$, so that spectral weight
at $\varepsilon_F$ comes from sites with $\epsilon_i^\prime = \pm
U/2$.  This remains approximately true when $t\neq0$ provided that
$W\gg t$.  Then, if site 1 contributes spectral weight at the Fermi
level,
\[
\varepsilon_F - \tilde \epsilon_1 \sim U,
\]
and Eq.~(\ref{sii2}) follows from Eq.~(\ref{sii}) provided
$|\varepsilon_F - \tilde \epsilon_2| > t^2/U $.  This condition will
only not be met when $\epsilon_2^\prime \approx -U/2$.  In other
words, the two cases not well described by Eq.~(\ref{sii2}) are (i)
$\epsilon_1^\prime \approx \epsilon_2^\prime$ and (ii)
$\epsilon_1^\prime \approx \epsilon_2^\prime + U$.

Certainly, in any random distributed set of site energies, both cases
are expected to occur for some fraction of sites on the lattice.
However, if the disorder potential is large, and the coordination
number of the lattice is low, then the probability of any given site
having a nearest neighbor satisfying either condition (i) or (ii) is
low, and the fraction of sites not described by Eq.~(\ref{sii2}) is
small.  It is interesting to note that the physical processes
neglected here are (i) formation of singlet correlations between
nearly degenerate sites and (ii) resonant exchange between sites in
which the double occupancy of site 1 is nearly degenerate with singlet
formation between sites 1 and 2.

Two further comments are warranted regarding our treatment of
$\Sigma_{11}$.  First, the simplifications made above assume that both
$U$ and $W$ are large.  This case is directly relevant to the current
work since it is the regime in which density of states anomalies are
observed.  However, we have found empirically through numerical
studies of small clusters that Eq.~(\ref{sii2}) is also a good
approximation when $U$ is small.  This point is illustrated in
Fig.~\ref{fig:twosite}(a).  Second, the term $t^\prime \Delta_1$ in
Eq.~(\ref{sii2}) is neglected in Eq.~(\ref{eq:sii3}).  In the clean
limit, $|\Delta| \lesssim 0.2$ and depends only weakly on
$U$.\cite{Avella2004} This term is small relative to the Hartree shift
$V\sum_j n_j$ and is therefore neglected.

We emphasize that the Hartree shift $V\sum_j n_j$ does not arise
naturally in our treatment of the Hubbard-$U$ interaction.  It must
therefore be undertstood to come directly from the nonlocal Coulomb
interaction.  This point is important because, as is shown in
Sec.~\ref{results}, the Hartree term makes a large contribution to the
ZBA at half-filling.

The off-diagonal self-energy is significantly harder to evaluate than
the diagonal term.  It requires knowledge of a quantity $p$ that
cannot be calculated self-consistently within the two-pole
approximation, although various approximate schemes exist for its
evaluation.\cite{Avella2004} However, we note from the definition of
$p$ that it measures exchange correlations between sites 1 and 2 and
plays a similar role to the exchange self-energy $-Vf_{12}$ defined in
Sec.~\ref{method}.  This suggests that, qualitatively, the exchange self-energy
may represent the exchange term from the nonlocal Coulomb
interaction.

Figure~\ref{fig:twosite} shows the DOS of a two-site system calculated
within the approximation (\ref{eq:Sapprox}).  The main effect of $V$
is to produce a level repulsion between states above and below the
Fermi energy, as illustrated in Fig.~\ref{fig:twosite}(b).  The
magnitude of the level repulsion decreases with increasing $U$, and is
unobservably small for $U=12t$ and $V<3t$.  As the figure shows, the
DOS is well-reproduced with $V=0$ when $U\ll W$ and $U\gg W$, but is
much less well reproduced when $U\approx W$.  As expected from the
analysis above, this can be corrected to some extent by a judicious
choice of $V$.

\subsection{Coherent Potential Approximation}
\label{cpa}

In this section, we describe an implementation of the coherent
potential approximation (CPA) that includes the effects of
interactions and disorder within an effective medium approximation.
As mentioned in the introduction, the CPA neglects spatial
correlations and therefore misses the physics of the ZBA.  It is
therefore useful as a point of comparison for our LDMFT
calculations. 

Our CPA implementation applies specifically to the HFA and
the HI approximation, and hence is denoted by HFA+CPA or HI+CPA as
appropriate, and reduces to these approximations in the limit
$W\rightarrow 0$.  The HFA+CPA and HI+CPA algorithms also reduce to
the usual CPA in the noninteracting $U\rightarrow 0$ limit.  Because
of the local nature of these approximations, the nonlocal interaction
$V$ has not been included.

For a particular lattice, we can calculate the local Green's function
of the disorder-averaged system:
\begin{equation}
G_\mathrm{loc}(\omega) = \frac{1}{N} \sum_{\bf k} \frac{1}{\omega - \epsilon_k - \Sigma(\omega)}
\label{eq:Gloc}
\end{equation}
In this equation, $\Sigma(\omega)$ is a self-energy that includes both
inelastic contributions from the local interaction and elastic
contributions from the disorder scattering.  On the first iteration of
the algorithm, $\Sigma(\omega)$ is guessed, and on later iterations it
is taken from the output of the previous iterations.  Next, we define
a Green's function
\begin{equation}
G_{\epsilon}(\omega) = \left [ [G_\mathrm{loc}(\omega)]^{-1} + \Sigma(\omega) - \epsilon - \Sigma_\epsilon(\omega) \right ]^{-1}
\label{eq:Ge}
\end{equation}
This is the local Green's function for a site with energy $\epsilon$
which is embedded in the effective medium.  The term
$\Sigma_\epsilon(\omega)$ is the {\em inelastic} self-energy for the
site, and must be determined self-consistently.  For both the HFA and
HI approximation, $\Sigma_\epsilon(\omega)$ depends on the local charge
density $n_\epsilon$.  Equation (\ref{eq:Ge}) can therefore be closed
by the relations
\begin{equation}
n_\epsilon = -\frac{2}{\pi} \int_{-\infty}^{\varepsilon_F} \mbox{Im }G_{\epsilon}(\omega) d\omega,
\label{eq:ne}
\end{equation}
and $\Sigma_\epsilon(\omega) = U \frac{n_\epsilon}{2}$ for the HFA or 
\begin{equation}
\Sigma_\epsilon(\omega) = U \frac{n_\epsilon}{2} +
\frac{U^2\frac{n_\epsilon}{2} (1- \frac{n_\epsilon}{2} )}{\omega -
\epsilon - U(1- \frac{n_\epsilon}{2})}
\label{eq:se}
\end{equation}
for the HI approximation.  Equations (\ref{eq:Ge})--(\ref{eq:se})
must be iterated to convergence for each value of $\epsilon$.

We then average $G_{\epsilon}(\omega)$ over site energies to get
\begin{equation}
G_\mathrm{av}(\omega) = \frac{1}{W} \int_{-W/2}^{W/2} d\epsilon
G_{\epsilon}(\omega),
\end{equation}
and a new self-energy is found via
\begin{equation}
\Sigma^\mathrm{new}(\omega) = [G_\mathrm{loc}(\omega)]^{-1} +
\Sigma(\omega) - [G_\mathrm{av}(\omega)]^{-1}
\end{equation}
The iteration cycle is now restarted at Eq.~(\ref{eq:Gloc}) with
$\Sigma^\mathrm{new}(\omega)$ taking the place of $\Sigma(\omega)$.
The iteration process is terminated when the difference between
$\Sigma^\mathrm{new}(\omega)$ and $\Sigma(\omega)$ is small.  When the
iteration cycle is complete, $G_\mathrm{loc}(\omega)$ is the
disorder-averaged Green's function of the interacting system.

\section{Results}
\label{results}
\subsection{Numerical Results, $V=0$}

\begin{figure}[tb]
\includegraphics[viewport= 0 00 480 360, width=\columnwidth, clip ]{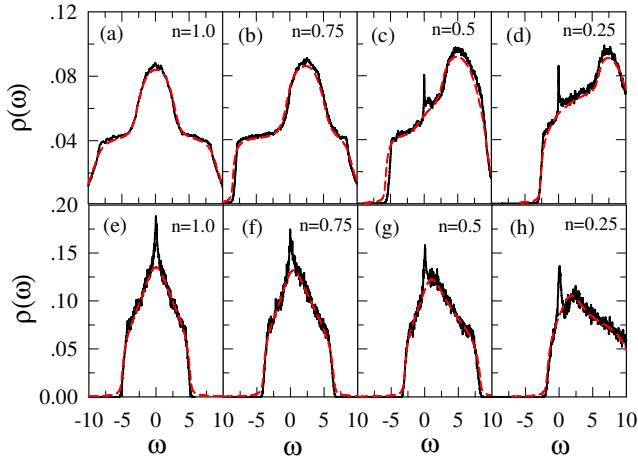}
\caption{(Color online) Evolution of the DOS with doping for a $
12\times 12$ lattice with $U=8t$, $W=12t$, $V=0$ and 1000-2000
impurity configurations. (a)-(d) Black solid lines and red dashed
lines represent the results of LDMFT and effective medium calculations
respectively. The Lorentz broadening is $\gamma=0.025$ throughout this
work.  (e)-(h) Corresponding plots for paramagnetic HFA calculations
are shown for the same parameters and 1000 impurity configurations.}
\label{fig:dosdop}
\end{figure}

We begin our discussion with the case $V=0$.  While it is not clear
that a system can be experimentally realized in which the off-diagonal
self-energy vanishes, this case is interesting because it provides a
relatively simple illustration of the role of strong correlations.
Throughout this section, strong correlation effects are identified by
comparisons between the HFA (which neglects correlations) and LDMFT.

The DOS is calculated from the self-consistently determined Green's
function via
\begin{equation}
\rho(\omega) = -\frac{1}{N\pi}\sum_i \mbox{Im }G_{ii}(\omega)
\end{equation}
where the rank-$N$ matrix ${\bf G}(\omega)$ is given by Eq.~(\ref{G}).
The evolution of $\rho(\omega)$ with doping is shown for $V=0$ in
Fig.~\ref{fig:dosdop}.  We have chosen $W=12t$, which corresponds to
$W=1.5D$ where $D = 8t$ is the bandwidth in the clean noninteracting
limit.  We have taken $U=8t$, which is large relative to $t$, but is
much less than the critical $U_c \approx W$ at which the Mott
transition takes place.  For comparison, we have shown results for
LDMFT and for the HI+CPA described in Sec.~\ref{cpa}.  These two
theories employ the same approximation for the interaction and differ
only in how they treat disorder: nonlocal spatial correlations between
impurity sites and the charge density are neglected in effective
medium approximations and are treated exactly in LDMFT.  The two
methods give quantitatively similar results, {\em except} for the ZBA
that emerges away from half filling in LDMFT but is absent in HI+CPA.

We note that the sign of the ZBA in Fig.~\ref{fig:dosdop} is
positive. This is different from the usual case discussed in the
literature, but is consistent with AA theory, where a negative ZBA
comes from the exchange self-energy while the Hartree self-energy
makes a weak positive correction to the DOS.  Since the Hubbard
interaction has a vanishing exchange self-energy, the expectation from
mean-field theory is for a positive ZBA when $V=0$.  This is
illustrated by the numerical HFA calculations shown in
Fig.~\ref{fig:dosdop}(e)-(h).

\begin{figure}[tb]
\includegraphics[width=\columnwidth]{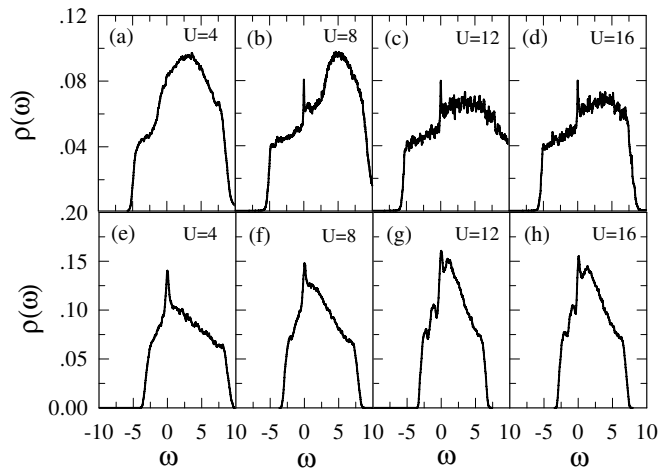}
\caption{Evolution of the DOS with $U$ at quarter-filling for $W=12t$,
$V=0$, and $8\times 8$ lattices with 1000 impurity configurations.
The upper panels (a)-(d) show LDMFT results while the lower panels
(e)-(h) show HFA results. } \label{fig:dosqfu}
\end{figure}
 
Although the sign of the ZBA is the same in LDMFT and HFA
calculations, its magnitude is different.  In particular, the peak at
$\varepsilon_F$ is finite at all doping levels in the HFA but is
absent near half-filling in the LDMFT calculations.  This difference
shows that strong correlations suppress the ZBA near half filling.
Results for quarter filling are shown in Fig.~\ref{fig:dosqfu} for a
range of $U$, where it can be seen that the ZBA grows with $U$ when
$U$ is small, but saturates when $U \gtrsim 8t$.  A more technical
discussion of these results is given in Sec.~\ref{analysis}, and we
briefly summarize the main ideas of this discussion here.

The main distinction between weakly and strongly correlated systems 
is that the local charge density $n_i$ is a
continuous variable in weakly correlated systems, but is restricted to
near-integer values in strongly correlated systems.  In the HFA, the energy
of an isolated site is $\omega_i = \epsilon_i + Un_i/2$.  For sites
with $\varepsilon_F - U < \epsilon_i < \varepsilon_F$, the
self-consistent equation for the charge density,
$n_i = 2f(\omega_i)$
[where $f(x)$ is the Fermi function], will be
satisfied at zero temperature by 
\[
\omega_i = \varepsilon_F, \qquad n_i =
2(\varepsilon_F-\epsilon_i)/U,
\]  
where the second equality comes from rearranging the expression for
$\omega_i$.  Since a macroscopic fraction of sites satisfy
$\varepsilon_F - U < \epsilon_i < \varepsilon_F$, a peak (i.e.\ a
positive ZBA) is expected in the DOS at the Fermi energy in the atomic
limit.  Numerical calculations (not shown) find that the peak
persists, but weakens, as $t/W$ grows.

In contrast, the poles in the spectral function for an isolated
strongly correlated site are at $\omega_i = \epsilon_i$ and $\omega_i
= \epsilon_i+U$.  Because of the rigidity of the relationship between
$\omega_i$ and $\epsilon_i$, the distribution of $\omega_i$ values
follows the distribution of $\epsilon_i$ and a vanishing fraction of
sites therefore have resonances at $\varepsilon_F$.  There is,
consequently, no ZBA when $t=0$; the ZBA in LDMFT calculations only
occurs when $t/W$ is nonzero.  The discussion in Sec.~\ref{analysis}
shows that the spectral weight in the ZBA is proportional to the
hybridization function $\Lambda_i(\varepsilon_F)$ between sites with
$\epsilon_i \approx \varepsilon_F$ or $\epsilon_i +U \approx
\varepsilon_F$ and the rest of the lattice.  The absence of a ZBA at
half-filling comes from the fact that these sites {\em decouple} from
the lattice, i.e.\ $\Lambda_i(\varepsilon_F) = 0$, when $\varepsilon_F
= U/2$.

Our analytical calculations in Sec.~\ref{analysis} also suggest that
$\Lambda_i(\varepsilon_F)$ is a strong function of both
$\varepsilon_F$ and $U$ provided $\varepsilon_F$ lies in the ``central
plateau" [by which we mean the broad peak in the DOS arising from the
overlap of upper and lower Hubbard bands; see, for example,
Fig.~\ref{fig:dosdop}(a)].  Outside the central plateau,
$\Lambda_i(\varepsilon_F)$ is a weak function of both $\varepsilon_F$
and $U$.  This is qualitatively consistent with the numerical results
 in Fig.~\ref{fig:dosqfu}, which show that the peak height
increases with $U$ for $U\lesssim 8t$ and saturates at larger $U$:  In
the limit $t/W \rightarrow 0$, it is easy to show that $\varepsilon_F$
lies in the central plateau for $U< W/2$.

\subsection{Numerical Results, $V\neq 0$}

\begin{figure}[tb]
\includegraphics[width=\columnwidth]{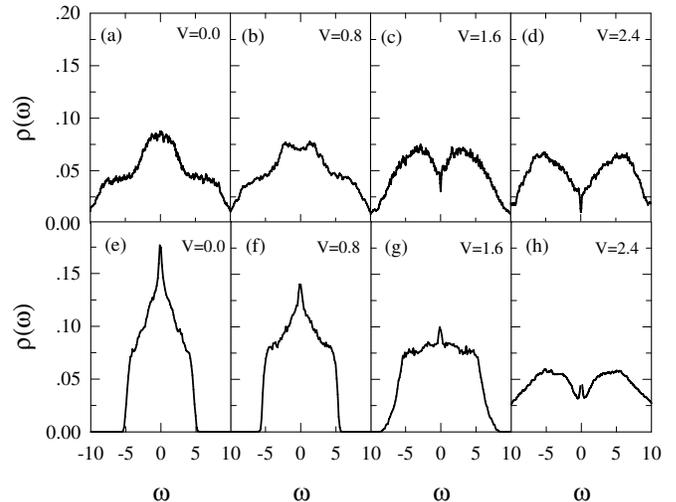}
\caption{Evolution of the DOS with $V$ at half filling for $U=8t$ and
$W=12t$. (a)-(d) LDMFT results for an $8\times 8$ lattice with
1000 impurity configurations.  (e)-(h) HFA results are for the
same parameters and $10\times 10$ lattices with 1000 impurity
configurations. }
\label{fig:dosV}
\end{figure}

We now consider the case $V \neq 0$.  As for $V=0$, the influence of
strong correlations is most visible at half filling.  For the purposes
of this discussion, the term ``ES-like behavior'' refers to a negative
Hartree contribution to the DOS, and the term ``AA-like behavior''
refers to a Hartree contribution which is positive and an exchange
contribution which is negative.

\begin{figure}
\includegraphics[width=\columnwidth]{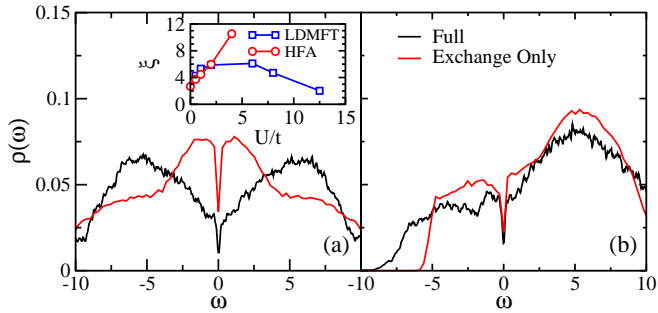}
\caption{(color online) Contributions to the zero bias anomaly for
$V=2.4t$ at (a) half filling and (b) quarter filling.  For the curve
labeled ``Full", both the exchange and Hartree self-energy
contributions from the nearest neighbor interaction are retained,
while for the curve labelled ``Exchange Only", the Hartree self-energy
is set to zero.  
The curves show that the ZBA comes primarily from the
Hartree contribution, and is therefore of the Efros-Shklovskii type.
The inset shows results for the localization length at half-filling
reproduced from a finite-size scaling analysis in
Ref.~\protect\onlinecite{Song2007} for $V=0$.  The short localization
length at large $U$ in LDMFT calculations, relative to HF
calculations, is consistent with the enhanced Coulomb gap in the LDMFT
calculations.}
\label{fig:cmpr}
\end{figure}

The DOS at half filling is shown in Fig.~\ref{fig:dosV} for $U=8t$ and
increasing $V$.  In both the LDMFT and HFA results ES-like and AA-like
behavior is present.  As discussed in \onlinecite{Vojta1998} there
appears to be a transition as a function of energy, with ES-like
behavior farther from the Fermi energy and AA-like behavior closer in.
Following the arguments of ES,\cite{Efros1975a} the average distance
between states  near the Fermi energy is very large.  In our case
the interaction has a finite range, and hence the ES behavior breaks
down very near the Fermi energy.  That the gap in Fig.~\ref{fig:dosV}
(d) is mostly ES-like, except right near the origin, can be seen in
Fig.~\ref{fig:cmpr} (a) where the full result and a result with the
Hartree term $V\sum_j n_j$ in Eq.~\ref{MFham} set to zero are
compared.  Note that the DOS does not satisfy the ES result
$\rho(\omega) \propto |\omega-\varepsilon_F|$ because the interaction
is short-range.  The most striking difference between the LDMFT and
HFA results is the more pronounced ES-like behavior in the
strongly-correlated case.  Strong correlations result in much less
screening of the disorder than in the mean-field treatment, hence
enhancing the disorder-driven ES-like behavior.  That the AA-like
behaviors in the LDMFT and HFA results differ in sign is not
surprising given the very different treatment of $U$ in the two cases.

\begin{figure}[tb]
\includegraphics[width=\columnwidth]{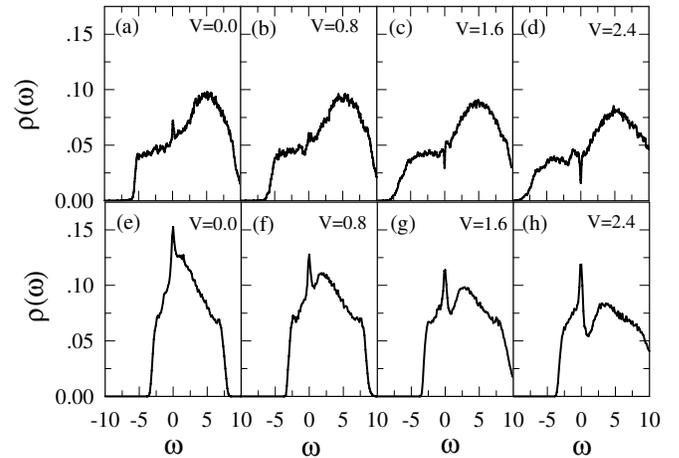}
\caption{Evolution of the DOS with $V$ at quarter filling with $W=12t$
and $U=8t$. Calculations are for an $8\times 8$ site lattice with more
than 1000 sample configurations for each parameter set.
Results are shown for (a)-(d) LDFMT and (e)-(h) HFA.}
\label{fig:dosqfv}
\end{figure}

The results at quarter filling are shown in Fig.~\ref{fig:dosqfv}.
The ZBA in the LDMFT results is less pronounced at quarter filling
than at half filling.  The narrow ZBA close to the Fermi energy
crosses over from positive to negative, consistent with increasing $V$
and hence increasing negative exchange contribution of the AA type.
As seen in Fig.~\ref{fig:dosqfv}(b), the Hartree self-energy modifies
the DOS over a large energy range, but does not produce a gap-like
feature at the Fermi energy.  Exact studies of small clusters, to be
reported elsewhere,\cite{Hongyi2009} suggest that the transition from
large to small ZBA occurs when $\epsilon_F$ is shifted outside the
central plateau described earlier. In the atomic limit, when
$\epsilon_F$ lies within the central plateau $n_i$ may have three
distinct values (0,1 or 2), whereas $n_i $ can only be 0 or 1 for
$\epsilon_F$ below the lower edge of the plateau.  In small clusters,
this reduction in the range of possible charge states for individual
sites directly results in a reduced ZBA.  In marked contrast to the 
strongly-correlated results, the HFA results show an AA-like peak 
and an ES-like dip, both of which grow with increasing $V$.

\begin{figure}[tb]
\includegraphics[viewport= 30 0 505 336, width=\columnwidth, clip ]{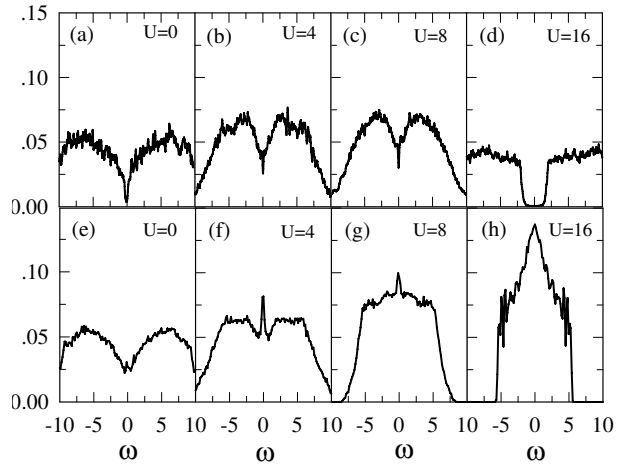}
\caption{Evolution of the DOS with $U$ at half-filling with $W=12t$
and $V=1.6t$. (a)-(d) LDMFT and (e)-(h) HFA calculations are shown.
Calculations are for an $8\times 8$ site lattice with more than 1000
sample configurations for each parameter set.}
\label{fig:dosqfU}
\end{figure}

Figure~\ref{fig:dosqfU} shows the variation of the DOS with $U$ for
$V=1.6t$.  For the somewhat artificial case of $U=0$ shown in
Fig. 6(a) and (e), the results differ slightly because the numerical
HFA and LDMFT routines converge differently.  Both obtain
charge-density-wave order frustrated by the disorder, but with
slightly shifted domain walls.  In the HFA case, increasing $U$ screens
the disorder, reducing the ES-like behavior.  The AA-like Hartree peak
is initially strengthened by $U$ but is reduced as the screening
increases.  At large $U$ [Fig.~\ref{fig:dosqfU}(h)], the DOS appraches
the clean-limit result.  In the LDMFT case, the screening produced by
$U$ initially weakens the ES-like behavior.  However, for larger
values of $U$, the screening in the strongly correlated case is much
less than that in mean field.  The localization length based on a
finite-size scaling analysis (calculated for $V=0$) is reproduced from
a previous work\cite{Song2007} in the inset of Fig.~\ref{fig:cmpr}.
While the localization length grows monotonically with $U$ in the HFA,
in the LDMFT it reaches a maximum at $U \approx 4t$ and decreases with
increasing $U$ thereafter.  Similarly, the ES-like behavior of the
LDMFT DOS is initially weakened, but is not lost as in the HFA 
%
 and saturates before the opening of the Mott gap.

We note that, in Fig.~\ref{fig:cmpr}, the exchange contribution to the
ZBA depends only {\em weakly} on doping.  We showed previously in
Sec.~\ref{connection} that the exchange self-energy arising from the
nonlocal interaction plays the same role as the exchange self-energy
arising from a perturbative treatment of intraorbital interaction.
Thus, the exchange-only curves shown in Fig.~\ref{fig:cmpr} should
behave the same qualitatively as an exact treatment of the $V=0$
Anderson-Hubbard model.  Such a study has been reported by Chiesa et
al.,\cite{Chiesa2008} where the ZBA was found to be independent of
doping for fillings $0.6 \leq n \leq 1$.  This is consistent with our
findings.

When the Hartree self-energy is included, our findings differ
substantially from Chiesa et al.  In particular, the doping-dependence
of the ZBA due to ES-like physics which we have reported here is not
found when the interaction is purely local.\cite{Chiesa2008} We
emphasize that the Hartree self-energy responsible for the ZBA is
physically distinct from the exchange contributions arising from the
Hubbard $U$ interaction, and can {\em only} occur when there is a
finite-ranged interaction.  In summary, the difference in results
between Chiesa et al.\ and us appears to stem from the difference in
the range of the electron-electron interactions.

\subsection{Analysis of strong correlation effects on the zero bias anomaly}
\label{analysis}

In this section, we discuss the origin of the ZBA in the $V=0$ case.
We begin with Eq.~(\ref{G}) for the Green's function.  In the large
disorder limit, it is possible to treat the hopping matrix element as
a perturbation.  When the matrix ${\bf t}$ is zero, ${\bf G}(\omega)$
decouples into a diagonal matrix describing an ensemble of isolated
atoms with Green's functions
\begin{equation}
G^0_{ii}(\omega) = \frac{1}{\omega - \epsilon_i - \Sigma^0_{ii}(\omega) },
\end{equation}
where the superscript zeros refer to the isolated atomic systems and $\Sigma^0_{ii}(\omega)$ is given exactly by Eq.~(\ref{eq:SE}).  The atomic Green's function, $G^0_{ii}(\omega)$, has poles at 
\begin{equation}
\omega^0_{i-}=\epsilon_i ; \quad
\omega^0_{i+} = \epsilon_i + U,
\end{equation}
with spectral weights 
\begin{equation}
Z^0_{i-}=1-\frac{n_{i}}{2}; 
\quad Z^0_{i+}=\frac{n_i}{2}.
\end{equation}  
The total density of states is found by averaging the imaginary part of $G^0_{ii}(\omega)$ over $\epsilon_i$ in the interval $-W/2 < \epsilon_i < W/2$.  Since the pole energies are linear functions of $\epsilon_i$, the total density of states is featureless at the Fermi energy.

Next, we use the fact that the diagonal matrix elements of Eq.~(\ref{G}) can be written in the
form
\begin{equation}
G_{ii}(\omega) = \frac{1}{\omega - \epsilon_i -\Lambda_i(\omega) - \Sigma_{i}(\omega)},
\label{eq:gii}
\end{equation}
where $\Lambda_i(\omega)$ is the hybridization function that describes the coupling of site $i$ to the rest of the lattice.  It is
\begin{eqnarray}
\Lambda_i(\omega) &=& \sum_{jk} t_{ij} t_{ki} G^{(i)}_{jk}(\omega) \nonumber \\
&\approx& t^2 \sum_j G^0_{jj}(\omega)
\label{eq:di}
\end{eqnarray}
where $t_{ij}$ are the matrix elements of ${\bf t}$ between sites $i$ and $j$, $G^{(i)}_{jk}(\omega)$
is the Green's function for the lattice with site $i$ removed, and the second line is the expansion of the first to $O(t^2)$.  Equation (\ref{eq:di}) applies formally to the limit that the localization length vanishes, but is qualitatively correct for $t \ll W$.  We note that $\Lambda_i(\omega)$ is a complex function of frequency for metallic systems, but is real with a discrete spectrum of simple poles for Anderson-localized systems as we have here.  
 
 Recalling Eq.~(\ref{eq:SE}), we solve for the poles of $G_{ii}(\omega)$  to $O(t^2)$: 
\begin{subequations}
\begin{eqnarray}
\omega_{i-} &=& \epsilon_i + \left (1-\frac{n_i}{2} \right ) \Lambda_i(\omega_{i-}),
\label{eq:polea}\\
\omega_{i+} &=& \epsilon_i + U + \frac{n_i}{2} \Lambda_i(\omega_{i+}),
\label{eq:poleb}
\end{eqnarray} 
\label{eq:pole}
\end{subequations} 
where we have assumed $\Lambda_i \ll U$.  In the approximation (\ref{eq:di}), $\Lambda_i(\omega_{i\pm})$
diverges when $\omega_{i\pm}$ is degenerate with any $\omega_{j\pm}$ for nearest neighbor site $j$.  This is an artifact of the approximation since any degeneracy between $i$ and $j$ is lifted by hybridization of the orbitals.  The poles of $\Lambda_i(\omega)$ must therefore differ from $\omega_{i\pm}$  by an energy $\gtrsim t$,  and we impose a cutoff $|\Lambda_i(\omega_{i\pm})| < t$.

The spectral weights $Z_{i\pm}$ of the poles (\ref{eq:pole}) are reduced by $O(t^2)$ from $Z^0_{i\pm}$ and the remaining spectral weight appears at new poles resulting from hybridization of site $i$ with the rest of the lattice.  
These poles play a role in suppressing the ZBA in the limit that the localization length becomes large, but are of secondary importance when $t \ll W$ as in the current discussion.

Equations (\ref{eq:pole}) contain the essential physics of the ZBA, which we summarize here before we go into the detailed calculations.  In both equations,
the local charge susceptibility $\chi_{ii} = -\partial n_i /\partial \epsilon_i $ is nonzero because of
the hybridization function $\Lambda_i(\omega)$.    
The main idea is that, because $\chi_{ii}$ is nonzero, sites with energies $\epsilon_i$  that are sufficiently close to $\varepsilon_F$ ($\varepsilon_F-U$) can adjust their filling $n_i$ such that $\omega_{i-}=\varepsilon_F$ ($\omega_{i+} = \varepsilon_F$).  The range of $\epsilon_i$ satisfying the criterion of ``sufficiently close" is set by $\Lambda_i(\varepsilon_F)$,
and the weight under the ZBA peak is therefore also set by $\Lambda_i(\varepsilon_F)$.  The suppression of the ZBA at half filling then follows from the fact that the disorder average
of $\Lambda_i(\varepsilon_F)$ is an antisymmetric function of $\varepsilon_F$.  

We consider sites with energies $\epsilon_i$ such that $\omega_{i\pm} = \varepsilon_F$.  The expression for $\omega_{i\pm}$ requires knowledge of the charge density $n_i$, which is given by 
$n_i/2 = \sum_\pm Z^0_{i\pm} f(\omega_{i\pm}) + O(t^2)$.  For sites with  $\omega_{i\pm} = \varepsilon_F$,
this reduces to [excepting terms of $O(t^2)$]
\begin{equation}
\frac{n_i}{2} = \left ( 1-\frac{n_i}{2} \right ) f(\varepsilon_F),  \quad (n_i<1) 
\label{eq:n1}
\end{equation}
for $\omega_{i-} = \varepsilon_F$ and
\begin{equation}
\frac{n_i}{2} = \left ( 1-\frac{n_i}{2} \right ) + \frac{n_i}{2}  f(\varepsilon_F),  \quad (n_i>1)
\end{equation}
for $\omega_{i+} = \varepsilon_F$.
At zero temperature, $0<f(\varepsilon_F) < 1$ and these equations are satisfied for a range of $\epsilon_i$.  Setting
$\omega_{i-} = \varepsilon_F$ in  Eq.~(\ref{eq:polea}) and applying the restriction $0 < n_i <1$, we generate the limits $\epsilon_L < \varepsilon_F - \epsilon_i < \epsilon_U$ on $\epsilon_i$, where
\begin{eqnarray*}
\epsilon_L &=& \min\left( \frac{\Lambda_i(\varepsilon_F)}{2}, \Lambda_i(\varepsilon_F)\right ), \\
\epsilon_U &=& \max\left( \frac{\Lambda_i(\varepsilon_F)}{2}, \Lambda_i(\varepsilon_F)\right ).
\end{eqnarray*}
In this range [rearranging Eq.~(\ref{eq:polea})]
\begin{equation}
1-\frac{n_i}{2} = \frac{\varepsilon_F - \epsilon_i}{\Lambda_i(\varepsilon_F)},
\end{equation}
and the density of states at $\varepsilon_F$ coming from sites with $\omega_{i-} = \varepsilon_F$ is
therefore
\begin{eqnarray}
\delta \rho_-(\omega \approx \varepsilon_F)& =& \frac{1}{W} \int_{\varepsilon_F -\epsilon_U}^{\varepsilon_F-\epsilon_L} d\epsilon_i Z_{i-} \delta(\omega-\varepsilon_F) \nonumber \\
&=& \frac 38 \frac{|\Lambda(\varepsilon_F)|}{W} \delta(\omega-\varepsilon_F) + O(t^4).
\end{eqnarray}
In this equation, $|\Lambda(\varepsilon_F)|$ is an average over $|\Lambda_i(\varepsilon_F)|$.  An identical
result can be found for sites with resonance energies $\omega_{i+} = \varepsilon_F$, so that the
total density of states near the Fermi energy is
\begin{equation}
\rho(\omega\approx \varepsilon_F) = \frac 34 \frac{|\Lambda(\varepsilon_F)|}{W} \delta(\omega-\varepsilon_F).
\label{eq:drho}
\end{equation}
Equation (\ref{eq:drho}) shows that the ZBA is a delta-function at zero temperature.  At finite temperatures $T$, the ZBA is a peak of width $\sim T$.   The spectral weight in the peak is proportional to the hybridization function $\Lambda(\varepsilon_F)$, and we next show how this depends on doping.

We consider a site $i$ in a lattice with coordination number $Z_c$, whose nearest neighbors have randomly chosen site energies.  At half filling ($\varepsilon_F = U/2$), the terms $G_{jj}^0(\varepsilon_F)$ in the sum in (\ref{eq:di}) are positive or negative with equal probability, and tend to cancel.  As the filling is reduced,  the probability that $G_{jj}^0(\varepsilon_F)$ is negative (positive) becomes larger (smaller).  We make a rough calculation that illustrates this behavior by replacing the sum over site index $j$ in (\ref{eq:di}) 
with an integral over $\epsilon_j$.  Thus
\begin{eqnarray}
\Lambda_i(\varepsilon_F) &\approx & \frac{t^2 Z_c}{W} \int_{-W/2}^{W/2} d\epsilon \left (
\frac{1-n_\epsilon/2}{\varepsilon_F-\epsilon}  + \frac{n_\epsilon/2}{\varepsilon_F-\epsilon-U} \right ),
\nonumber 
\end{eqnarray}
where $n_\epsilon$ is the charge density for sites with site-energy $\epsilon$.  
Recalling our constraint $|\Lambda_i(\varepsilon_F)|\lesssim t$, we introduce cutoffs near the poles of the integrand.  Noting that 
\begin{equation*}
n_\epsilon = \left \{\begin{array}{ll} 2,& \epsilon < \varepsilon_F - U\\
                                                               1, & \varepsilon_F - U < \epsilon < \varepsilon_F\\
                                                               0,&\epsilon > \varepsilon_F
\end{array}\right .                                                             
\end{equation*}
we get
\begin{subequations}
\begin{equation}
\Lambda_i(\varepsilon_F) \approx \frac{t^2 Z_c}{W}  \ln \left ( \frac{\frac W2 + \varepsilon_F - U}{\frac W2 - \varepsilon_F} \right ), 
\label{eq:d1}
\end{equation}
for $U-\frac W2  < \varepsilon_F <\frac W2 $ and
\begin{eqnarray}
\Lambda_i(\varepsilon_F) &\approx&  \frac{t^2 Z_c}{W}  \left [ 
\ln \left ( \frac{ \frac W2 + \varepsilon_F}{\frac W2 - \varepsilon_F} \right ) \right .
+ \frac 12 \ln \left ( \frac{ U - \frac W2 - \varepsilon_F}{\frac W2 + \varepsilon_F} \right )  \nonumber \\
&&\left  . 
+ \frac 12 \ln \left ( \frac tU \right ) 
\right ]
\label{eq:d2}
\end{eqnarray}
\label{eq:ds}
\end{subequations}
for $-\frac W2 < \varepsilon_F < U-\frac W2$.    The logarithmic divergences in Eqs.~(\ref{eq:ds}) are artificial and must be cut off whenever any numerator or denominator has a magnitude smaller than $t$.   
Equation (\ref{eq:d1}) applies when the Fermi level sits in the central plateau, and shows that $\Lambda_i(\varepsilon_F)$ is antisymmetric about half-filling (i.e.\ $\varepsilon_F = U/2$), and grows linearly away from half-filling.  Outside of the central plateau, $\Lambda_i(\varepsilon_F)$ is
a weak function of  $\varepsilon_F$.  

In order to compare with Fig.~\ref{fig:dosqfu}, we 
evaluate Eq.~(\ref{eq:d2}) at quarter filling, which for small $t$ corresponds to
\[
\varepsilon_F \approx \left \{ \begin{array}{ll} 
\frac{U}{2}-\frac{W}{4}, & U < \frac{W}{2} \\
0, & U > \frac{W}{2} 
\end{array} \right.
\]
Then Eq.~(\ref{eq:d2}) gives
\begin{subequations}
\begin{equation}
|\Lambda_i(\varepsilon_F)| \approx  \frac{t^2 Z_c}{W}  \left[ \ln \left ( \frac{\sqrt{(\frac W2)^2-U^2}}{\frac 32W-U} \right )
- \frac 12 \ln \left ( \frac{U}{t} \right ) 
\right ],
\end{equation}
for $U<\frac W2$, and
\begin{equation}
|\Lambda_i(\varepsilon_F)| \approx  \frac{t^2 Z_c}{2W}  \left[ \ln \left ( \frac{2U- W}{W} \right )
- \ln \left ( \frac{U}{t} \right ) 
\right ],
\end{equation}
\label{eq:peak}
\end{subequations}
for $U>\frac W2$.  In Eqs.~(\ref{eq:peak}),
$|\Lambda_i(\varepsilon_F)|$ grows linearly with $U$ for small $U$
(recall that there is a cutoff such that $\ln(U/t) \rightarrow
\ln(t/t)$ when $U<t$), and saturates at a finite value when $U\gg
\frac W2$.  Both these results, and the results at half filling in
Eq.~(\ref{eq:d1}) are qualitatively consistent with the numerical
results shown in Figs.~\ref{fig:dosdop} and \ref{fig:dosqfu}.

\section{Conclusions}
\label{conclusions}
We have studied the effects of strong correlations on the zero bias 
anomaly in the density of states for disordered interacting systems. 
Our results show significant doping dependence.  When only local 
interactions are included, a positive ZBA is suppressed by strong 
correlations at half filling due to cancelation in the hybridization 
function responsible for the peak.  When nearest-neighbor 
interactions are included (simultaneously producing a more accurate 
treatment of the Hubbard $U$ term), strong correlations modify the 
mean field results at both half and quarter filling.  In particular, 
at half filling ES-like behavior is enhanced due to reduced screening 
in the strongly correlated system, whereas at quarter filling the ZBA 
is smaller reflecting the doping dependence of the ES physics.

\section*{Acknowledgments}
We acknowledge financial support from NSERC of Canada, Canada Foundation for Innovation, Ontario
Innovation Trust, and the DFG (SFB 484).  Some calculations were performed on the High Performance Computing Virtual Laboratory.


\begin{thebibliography}{20}
\bibitem{caveat} Note that in cases where the self-energy can be expressed analytically in terms of
$G_{ii}(\omega)$ or in terms of expectation values derived from $G_{ii}(\omega)$, it is possible to eliminate the step of explicitly constructing the Anderson impurity problem.  This presents a considerable simplification and in the current case reduces the problem to one which is equivalent to solving a combination of Hartree-Fock and Hubbard-I equations in real space on a disordered lattice.
\end{thebibliography}
\end{document}